\documentclass[twocolumn]{aastex62}

\usepackage{amsmath}
\usepackage{units}
\usepackage{hyperref}

\newcommand{\MESA}{\texttt{MESA}}

\newcommand{\gcc}{\ensuremath{\mathrm{g\,cm^{-3}}}} 
\newcommand{\Ye}{\ensuremath{Y_{\mathrm{e}}}} 
\newcommand{\Msun}{\ensuremath{\mathrm{M}_{\sun}}} 

\newcommand{\BV}{Brunt-V\"{a}is\"{a}l\"{a}}

\newcommand{\abar}{\bar{A}}
\newcommand{\zbar}{\bar{Z}}
\newcommand{\chiT}{\chi_{T}}
\newcommand{\chir}{\chi_{\rho}}
\newcommand{\mue}{\mu_{\mathrm{e}}}

\newcommand{\grada}{\nabla_{\mathrm{ad}}}
\newcommand{\gradT}{\nabla_{T}}

\newcommand{\ppll}[2]{\left(\frac{\partial\ln #1}{\partial\ln
      #2}\right)_{\rho,T,\{X_{j\neq i}\}}}
\newcommand{\ddl}[2]{\frac{d \ln #1}{d \ln #2}}

\newcommand{\vect}[1]{\mathbf{#1}}
\newcommand{\unitvec}[1]{\vect{e}_#1}
\renewcommand{\Pr}{\textrm{Pr}}

\newcommand{\Ra}{\textrm{Ra}}
\newcommand{\abs}[1]{|#1|}

\newcommand{\num}{\textrm{Nu}_{\mu}}

\received{}
\revised{}
\accepted{}

\submitjournal{ApJ}

\shorttitle{Hybrid mixing}
\shortauthors{Schwab \& Garaud}
\watermark{}

\begin{document}


\title{Mixing via Thermocompositional Convection in Hybrid C/O/Ne White Dwarfs}

\author[0000-0002-4870-8855]{Josiah Schwab}
\altaffiliation{Hubble Fellow}
\affiliation{Department of Astronomy and Astrophysics, University of California, Santa Cruz, CA 95064, USA}
\correspondingauthor{Josiah Schwab}
\email{jwschwab@ucsc.edu}

\author{Pascale Garaud}
\affiliation{Department of Applied Mathematics, Baskin School of Engineering, University of California, Santa Cruz, CA 95064, USA}

\begin{abstract}
  Convective overshooting in super asymptotic giant branch stars has
  been suggested to lead to the formation of hybrid white dwarfs with
  carbon-oxygen cores and oxygen-neon mantles.  As the white dwarf
  cools, this core-mantle configuration becomes convectively unstable
  and should mix.  This mixing has been previously studied using
  stellar evolution calculations, but these made the approximation
  that convection did not affect the temperature profile of the mixed
  region.  In this work, we perform direct numerical simulations of an
  idealized problem representing the core-mantle interface of the
  hybrid white dwarf.  We demonstrate that, while the resulting
  structure within the convection zone is somewhat different than what
  is assumed in the stellar evolution calculations, the two approaches
  yield similar results for the size and growth of the mixed region.
  These hybrid white dwarfs have been invoked as progenitors of various
  peculiar thermonuclear supernovae.  This lends further support to
  the idea that if these hybrid white dwarfs form then they should be
  fully mixed by the time of explosion.  These effects should be included
  in the progenitor evolution in order to more accurately characterize
  the signatures of these events.
\end{abstract}


\section{Introduction}
\label{sec:intro}

Oxygen-neon white dwarfs (ONe WDs) are formed when off-center carbon
burning is ignited in an asymptotic giant branch (AGB) star and a
convectively-bounded, carbon-burning deflagration wave (the ``carbon
flame'') propagates to the center of the star
\citep[e.g.,][]{GarciaBerro1994, Siess2006, Farmer2015}.  Mixing at
the lower convective boundary can lead to the quenching of the carbon
flame \citep{Siess2009, Denissenkov2013b}.  If this occurs, a
``hybrid'' WD consisting of an unburned carbon-oxygen (CO) core
surrounded by an ONe mantle is formed.

\citet{Denissenkov2013b} use stellar evolution calculations and a
commonly used exponential overshooting parameterization
\citep{Herwig2000} to show that this flame quenching occurs in AGB
star models for even small amounts of overshoot \citep[see also][]{Chen2014c}.
However, using
direct numerical simulations of an idealized carbon flame model,
\citet{Lecoanet2016c} conclude that buoyancy prevents the convective
plumes from penetrating deeply into this interface and find
insufficient mixing occurs to disrupt the flame \citep[see
also][]{Lattanzio2017}.  Nevertheless, the possibility that such
hybrid WDs form has generated substantial interest, as such objects
may be promising progenitors of peculiar thermonuclear supernovae
\citep{Meng2014, Denissenkov2015, Kromer2015, Bravo2016}.

Existing work exploring the observational consequences of the
explosion of these hybrid WDs has made the assumption that the CO/ONe
core-mantle structure of the WD is intact at the onset of the carbon
simmering phase or at the time of explosion.  However,
\citet{Brooks2017a} point out that the core-mantle interface will
become convectively unstable as the WD cools.  Indeed, weak reactions during
carbon burning lower the electron fraction (\Ye) of the ONe mantle
relative to the CO core.  The configuration is initially stable
because the ONe ash is hotter than the unburned CO, but as global cooling and
thermal conduction reduce both the temperature and the temperature
gradient in the WD, the unstable composition gradient will ultimately
drive thermocompositional convection that can destroy the sharp
core-mantle interface.

\citet{Brooks2017a} use a stellar evolution code, Modules for
Experiments in Stellar Astrophysics \citep[\MESA; ][]{Paxton2011,
  Paxton2013, Paxton2015, Paxton2018}, to model cooling hybrid WDs and
demonstrate the efficiency of this mixing process.  Numerical
considerations caused them to make the assumption that only
the composition gradients and not the temperature gradients were modified in convectively unstable
regions.  In this work, we use direct numerical simulations of thermocompositional convection, in both double-diffusive and overturning regimes, to study an idealized version of the situation
present in these cooling hybrid WDs.  We find that while the
assumption of an unmodified temperature gradient is a
poor one, the global timescale for compositional mixing remains
analogous to that found by \citet{Brooks2017a}.\footnote{
We find that the convective regions grow and mix material on
a timescale set by the selected cooling timescale in our simulations.  
Likewise, the mixing prescription assumed in the stellar evolution calculations
of \citet{Brooks2017a} led to the mixing of the stellar model on its
cooling time.}
Therefore, we
strengthen the argument that even if hybrid WDs do form, most should
be well-mixed at the time of explosion.

In Section~\ref{sec:problem} we review what is known of the evolution
of cooling hybrid WDs.  In Section~\ref{sec:numerics} we outline the
equations being solved in our numerical simulations and devise an
idealized problem that captures the key features of the cooling hybrid
WD.  In Section~\ref{sec:comparions} we compare the results of our
simulations to models that adopt the approach taken by
\citet{Brooks2017a}.  In Section~\ref{sec:conclusions} we summarize
and conclude.

\section{Cooling Hybrid WDs}
\label{sec:problem}

\begin{figure}
\includegraphics[width=\columnwidth]{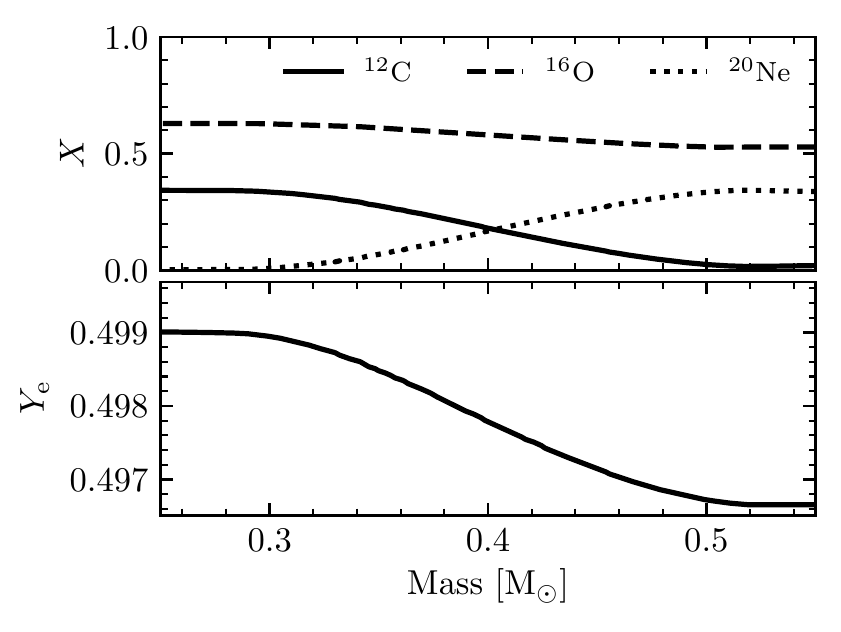}
  \caption{Composition of the example \MESA\ hybrid WD model around
    the region where mixing will begin.  The upper panel shows the
    mass fraction of the main isotopes, illustrating the transition
    from the unburned CO core to the ONe mantle.  The lower panel
    shows the electron fraction; the material processed by carbon
    burning has a lower \Ye.}
  \label{fig:mesa-hybrid-composition}
\end{figure}

We briefly review the evolution that leads these hybrid WDs to become
unstable to convective mixing.  As an illustrative case,
Figure~\ref{fig:mesa-hybrid-composition} shows the composition of the
hybrid WD model from \citet{Brooks2017a} that has a total
mass of 1.09 \Msun\ and a 0.4 \Msun\ CO core.  The hybrid WD has lower \Ye\ material
above higher \Ye\ material.

The Ledoux criterion for convective instability is \mbox{$N^2 < 0$}, where $N$ is
the total \BV\ frequency.  We can write
\begin{equation}
  N^2 = \frac{g^2 \rho}{P}\frac{\chiT}{\chir}
  \left(\grada-\gradT+B\right),
  \label{eq:brunt_B}
\end{equation}
where $\gradT$ and $\grada$ are the actual and adiabatic temperature
gradients, $g$ is the local gravitational acceleration, $\rho$ and $P$
are the density and total pressure, and
$\chi_T=\partial\ln P/\partial\ln T$ and
$\chi_\rho=\partial\ln P/\partial\ln \rho$ are the compressibilities.
The effect of composition gradients is included in the $B$ term
\citep[Equation~6 in][]{Paxton2013},
\begin{equation}
  B = -\frac{1}{\chiT} \sum_{i=1}^{N-1} \ppll{P}{X_i}\ddl{X_i}{P},
  \label{bsum}
\end{equation}
where $X_i$ is the mass fraction of species $i$ and the sum runs over
all but one of the species (reflecting the constraint that
$\sum_i X_i = 1$).  In practice, the influence of the composition can
be mostly specified by the mean ion weight, $\abar$, and the mean ion
charge, $\zbar$ \citep[see Equation~5 in][]{Brooks2017a}.
In the perhaps more familiar case with an equation of state $P = P(\rho,T,\mu)$,
the $B$ term is usually written as $B = (\phi /\delta)\nabla_{\mu}$, 
where $\delta = -(\partial \ln \rho/\partial \ln T)$,
$\phi = (\partial \ln \rho / \partial \ln \mu)$, and
$\nabla_{\mu} = (d \ln \mu / d\ln P)_s$.

\begin{figure}
  \includegraphics[width=\columnwidth]{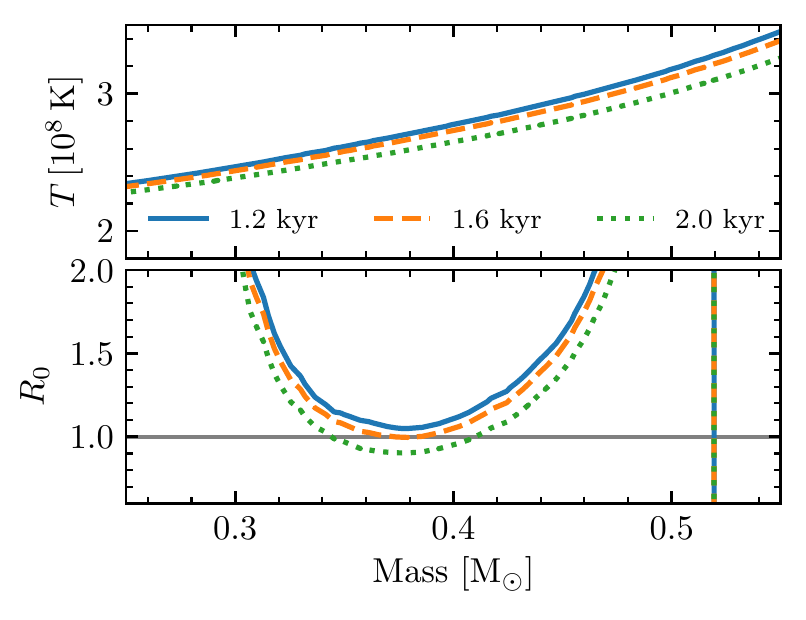}
  \caption{Cooling of the example hybrid WD.  The upper panel shows
    the temperature; the lower panel shows the density ratio.  For
    illustrative purposes, mixing is not allowed to occur (so the
    composition profile remains fixed at that shown in
    Figure~\ref{fig:mesa-hybrid-composition}).  The model becomes
    unstable to overturning convection after 1.6 kyr of cooling.}
  \label{fig:mesa-hybrid-cooling}
\end{figure}

We can then write the criterion for instability to overturning
convection in terms of the density ratio, $R_0$, as
\begin{equation}
R_0 \equiv \frac{\gradT - \grada}{B} < 1
\label{eq:R0}
\end{equation}
Figure~\ref{fig:mesa-hybrid-cooling} shows the temperature and density
ratio in the cooling hybrid WD model.  This model is evolved forward
in time with convective mixing disabled.  As the hybrid WD cools, the
temperature decreases and the temperature gradient moves towards
isothermal.  The decreasing temperature causes the magnitude of $B$ to
increase because $\chiT \propto T$ for degenerate
material.  The increasing (becoming less negative) value of $\gradT$
means that both the numerator and denominator in Equation~\eqref{eq:R0}
move towards overturning convection, which sets in after only 1.6 kyr
of cooling.  Note that from the beginning, the hybrid WD is unstable
to double-diffusive convection.  Even before the flame quenches, the
region with the composition gradient is unstable to fingering
(thermohaline) convection \citep{Siess2009}.  However, the timescale
for this double-diffusive mixing is sufficiently long that it does not
have a significant effect during the flame propagation
\citep{Denissenkov2013b} and thus also not in this brief initial
cooling period.

When modeling these cooling WDs in \MESA, \citet{Brooks2017a}
encounter numerical difficulties when using the standard
implementation of mixing length theory (MLT).  Cells alternate between being
convective and non-convective, meaning the temperature gradient
($\gradT$) at cell faces switches between the adiabatic temperature
gradient ($\nabla_{\rm ad}$) and the radiative temperature gradient
($\nabla_{\rm rad}$).  This causes problems for the iterative solver
that advances the stellar evolution model in time.  To circumvent
this, they assume $\gradT$ is $\nabla_{\rm rad}$ in convective regions
instead of $\nabla_{\rm ad}$.  This assumes that the energy transport
via convection does not significantly modify the thermal evolution of
the WD.  The physical reason for this numerical difficulty appears to
be because the timescale for mixing from MLT is much shorter than the
cooling time (and thus the timestep taken in the stellar evolution
models).

\section{Direct Numerical Simulations}
\label{sec:numerics}

In what follows, we construct the simplest possible model that mimics the onset of thermocompositional convection 
as the star cools, and study the resulting mixing processes that take place by solving the full nonlinear hydrodynamic equations numerically. 
We now describe the model setup and equations that
we solve (Section \ref{sec:equations}), the initial conditions of our
idealized problem (Section~\ref{sec:initial}), discuss parameter selection (Section \ref{sec:parameters}), and then show results
of the calculations (Section~\ref{sec:paddi}).

\subsection{Equations}
\label{sec:equations}

We focus on a small region located in the vicinity of the CO/ONe core-mantle interface. We 
use a 2D Cartesian coordinate system ($x$, $z$), with gravity
given by $\vect{g}=-g\unitvec{z}$. We are restricted to 2D simulations because of the vast range of scales that need to be accounted for numerically -- from the tiniest scale associated with double-diffusive fingering, to the global scale associated with convection. This assumption is discussed in more detail in Section~\ref{sec:conclusions}. We assume for simplicity that the temperature has
a linear background state and the mean molecular weight profile has a
constant background state,
\begin{align}
  T_0(z,t) &= T_{\mathrm{m}} + zT_{0z}(t) \\
  \mu_0(z,t) &= \mu_{\mathrm{m}}~.
\end{align}
The values $T_{\rm m}$ and $\mu_{\rm m}$ given here can be interpreted as the mean temperature and mean molecular weight per free electron of the pure CO core. The unstable compositional profile due to the CO/ONe transition, as well as perturbations to that state, will be added as initial conditions to drive the thermocompositional instability (see Section \ref{sec:initial}). While $T_{\rm m}$ could be time-dependent in a real WD, that time-dependence does not affect the hydrodynamics of the system much, so we neglect it here as a first approximation. By contrast, the
temperature {\it gradient} $T_{0z}$ is assumed to be time-dependent to model the effect of cooling, as in 
\begin{equation}
  T_{0z}(t) = T_{0z}(t=0) \exp(-t/\tau_{\rm cool})~,
  \label{eq:Tgradient}
\end{equation}
where $\tau_{\rm cool}$ is a characteristic global cooling timescale. This is a simple way of accounting for the gradual flattening of the temperature profile towards an isotherm, and the associated decay of the stabilizing temperature stratification, as the star cools. 


The total pressure in the WD material can be approximated as a
zero-temperature electron gas plus an ideal gas of ions,
\begin{equation}
  P = P_{\rm e} + P_{\rm ion} = K \left(\frac{\rho}{\mue}\right)^\gamma + \frac{\rho R T}{\abar},
  \label{eq:eos}
\end{equation}
where recall that $\mue$, the mean molecular weight per free electron, is defined as $\mue \equiv 1/\Ye = \abar/\zbar$.
This assumes a simple polytropic approximation for the electron pressure, with a constant $K$ and an adiabatic index $\gamma$ which would fall in the range 5/3 (non-relativistic electrons) to 4/3 (relativistic electrons).
At the conditions of
interest, the ion pressure contributes at the percent level.
Finite-temperature corrections to the electrons and non-ideal
contributions from the ions, both which are neglected in the above
expression, also enter at the percent level.
In what follows, we use the Boussinesq approximation for gases \citep{SpiegelVeronis1960}, in which fluid parcels are assumed to remain in pressure equilibrium with their surroundings at all times. 
Linearizing the equation of state (Equation~\ref{eq:eos}), and  setting pressure perturbations $\delta P$ to zero, we obtain an explicit expression for density perturbations $\delta  \rho$ in terms of temperature perturbations $\delta T$ and perturbations to the mean molecular weight per free electron $\delta \mu_e$: 
\begin{equation}
  \frac{\delta \rho}{\rho} \equiv -\alpha \delta T + \beta \delta \mue \approx -\left(\frac{P_{\rm ion}}{\gamma P}\right) \frac{\delta T}{T} + \frac{\delta \mue}{\mue},
\end{equation}
where we have used the fact the ion pressure is a small fraction of
the total in order to neglect its composition dependence.  This
expression looks like the expression for an ideal gas, but with a
thermal expansion coefficient, $\alpha$, that is suppressed by a
factor $P_{\rm ion}/(\gamma P) \sim 0.01$, and the mean molecular
weight per free electron $\mue$ playing the analogous role to the
total mean molecular weight.  Recognizing this analogy, for
notational convenience we will drop the subscript ``e''
and elide the ``per free electron'' in what follows.

Then, the equations for the perturbations around the mean state
($T = T_0 + \widetilde{T}$, $\mu = \mu_0 + \widetilde{\mu}$) satisfy
\begin{align}
\frac{\partial{\vect{u}}}{\partial{t}}+\vect{u}\cdot\nabla\vect{u}=-\frac{\nabla{p}}{\rho_{\mathrm{m}}}+g(\alpha{\widetilde{T}}-\beta\widetilde{\mu})\unitvec{z}+\nu\nabla^2\vect{u}&,  \\
  \frac{\partial{\widetilde{T}}}{\partial{t}}+\vect{u}\cdot\nabla{\widetilde{T}}+w\left(T_{0z}(t)-T_{\textrm{ad},z}\right)= \kappa_T\nabla^2\widetilde{T}&, \\
  \frac{\partial\widetilde{\mu}}{\partial{t}}+\vect{u}\cdot\nabla\widetilde{\mu}= \kappa_\mu\nabla^2{\widetilde{\mu}}&, \\
  \nabla \cdot \vect{u} = 0 
\end{align}
where $\vect{u}=(u,v,w)$ is the fluid velocity, $p$ is the pressure, $\rho_{\rm m}$ is the mean density of the region,
$\nu$ is the kinematic viscosity, $\kappa_T$ is the thermal
diffusivity, $\kappa_\mu$ is the compositional diffusivity, and
$T_{\textrm{ad},z} = -g/c_P$ is the background adiabatic temperature
gradient.  The relevant compositional diffusivity is the diffusivity of the ions,
as charge neutrality means that the mean molecular weight per free electron can only change 
due to the transport of ions.  The relevant thermal diffusivity is dominated by electron transport.
The thermal expansion and compositional contraction
coefficients are denoted $\alpha$ and $\beta$ respectively.  All of
these quantities are assumed to be constant for simplicity. This assumption can of course affect detailed predictions of the model, but not its general conclusions (see the discussion in Section~\ref{sec:conclusions}).

We then non-dimensionalize the equations. We take as the characteristic length
scale $[l] = d=(\kappa_T\nu/(g\alpha \abs{T_{\textrm{ad},z}}))^{1/4}$.
The
characteristic time, temperature, and composition scales are
$[t]=d^2/\kappa_T$, $[T]=d |T_{\textrm{ad},z}|$, and
$[\mu]=(\alpha/\beta)d|T_{\textrm{ad},z}|$ respectively. We define the
Prandtl number to be $\Pr\equiv\nu/\kappa_T$ and the diffusivity
ratio to be $\tau\equiv\kappa_{\mu}/\kappa_T$.
The non-dimensionalized Boussinesq equations are
\begin{align}
  \label{eq:governing-start}
\frac{1}{\Pr}\left(\frac{\partial{\vect{\widehat{u}}}}{\partial{t}}+\vect{\widehat{u}}\cdot\nabla\vect{\widehat{u}}\right)=-\nabla{\widehat{p}}+(\widehat{T}-\widehat{\mu})\unitvec{z}+\nabla^2{\widehat{u}}&, \\
\frac{\partial{\widehat{T}}}{\partial{t}}+\vect{\widehat{u}}\cdot\nabla{\widehat{T}}+w(s e^{-t/\tau_{\rm cool}}+1)=\nabla^2\widehat{T}&, \\
\frac{\partial{\widehat{\mu}}}{\partial{t}}+\vect{\widehat{u}}\cdot\nabla\widehat{\mu}=\tau\nabla^2\widehat{\mu}&, \\
  \nabla \cdot \widehat{\vect{u}} = 0, 
\label{eq:governing-end}
\end{align}
where $s = T_{0z}(t=0)/|T_{{\rm ad},z}|$. Note that the vertical potential density gradient is given by 
\begin{equation}
\frac{d \widehat \rho}{dz} - \frac{d \rho_{\rm ad}}{dz} = - \left( s e^{-t/\tau_{\rm cool}}+1 + \frac{d \widehat T}{dz}   \right) + \frac{d\widehat  \mu}{dz}
\end{equation}
where $\frac{d \rho_{\rm ad}}{dz} =1$ in the system of units selected.
The Ledoux threshold for convective instability then conveniently reduces to requiring that the potential density gradient be zero.

\subsection{Initial conditions}
\label{sec:initial}
In what follows, we solve these equations using the pseudo-spectral double-diffusive
convection code PADDI.
This code is described in \citet{Traxleral2011}, and has been used extensively both in the geophysical
\citep{Stellmachal2011, Garaud2015b, Reali2017, Brown2019} and astrophysical
\citep[e.g.,][and many others]{Rosenblum2011a, Mirouh2012, Brown2013b, Garaud2016a} literature to date.
PADDI uses periodic
boundary conditions in all directions. As such, the initial conditions are constrained to be periodic as well, and must be chosen carefully to avoid any effect of the boundary conditions on the dynamics of the convective region that develops near the CO/ONe core-mantle interface.
We have verified a posteriori for each run presented here that the domain boundaries are far from the convective region and have no influence over it (see later on Figure~\ref{fig:Profiles} for instance).

We select an initial mean molecular weight perturbation profile $\widehat{\mu}$ that is zero in the lower part of the domain (so the total mean molecular weight tends to $\mu_m$, which is that of the CO core), and increases relatively sharply outward at some specified height to model the presence of the CO/ONe core-mantle interface. Close to the upper boundary, the mean molecular weight perturbation is assumed to drop to zero again to satisfy the periodicity requirement. This upper transition does not affect the dynamics of the system much, as long as it is sufficiently far from the core-mantle boundary (see below for more on this topic).  
Since the background stable 
temperature gradient gradually decreases with time, our initial conditions for $\widehat{\mu}$ should also be selected to model a state
that is not initially unstable to overturning convection, but that
becomes so as the simulation proceeds.

To this end, our initial condition for the mean molecular weight is %
\begin{equation}
  \label{eq:mu-init}
  \widehat{\mu}_{\rm init}(z) = \frac{\Delta \mu}{2} \left[\tanh\left(\frac{z-z_I}{\sigma_I}\right) - \tanh\left(\frac{z-z_T}{\sigma_T}\right)  \right]  ~.
\end{equation}
The first $\tanh$ in $\widehat{\mu}_{\rm init}$ represents the core-mantle interface centered around $z_I$, with an initial
destabilizing composition gradient 
\begin{equation}
  \frac{d\widehat{\mu}_{\rm init}}{dz}(z = z_I) = \frac{\Delta \mu}{2 \sigma_I}~.
\end{equation}
The size $\Delta \mu$ and width $\sigma_I$ of the mean molecular weight jump are two of the input parameters of the system, and must be carefully selected to achieve the desired parameter regime (see Section \ref{sec:parameters}). The second $\tanh$ in $\widehat{\mu}_{\rm init}$ is required so that we satisfy the
periodic boundary conditions required by our code. It is centered a little below the top of the computational domain, and the transition is selected to be very sharp to offer a strongly stabilizing buffer which prevents any remaining fluid motions in that region from moving through the top boundary.
Our domain size will be selected to be sufficiently large such that the turbulent motions of interest do not reach this transition region.

The initial conditions for the velocity fields are selected to be zero.
A small amount of noise%
\footnote{A field has a different random number added to it at each grid point in the domain.  These numbers are distributed uniformly between [-1,1], multiplied by some small amplitude.}
is added to the initial temperature and mean molecular weight
profiles to drive the instability.
We selected the initial amplitude of the random perturbations
in the mean molecular weight to be $10^{-3}$ and in the temperature to be $10^{-1}$.
This is sufficiently
small that the initial development of the instability is in the
exponentially-growing, linear phase.  It is sufficiently
large that we reach the more interesting nonlinear stage of the
simulation on a timescale much shorter than that over which the
gradients imposed in the initial condition would diffuse away.
With these conditions met, the results will be independent of the initial choice of
perturbation amplitude.

\subsection{Parameter selection}
\label{sec:parameters}

The non-dimensional equations and initial conditions presented earlier contain 11 parameters, including some related to (1) the geometry of the system, characterized by the global domain dimensions $(L_x,L_z)$ and the respective heights and widths of the transitions in the initial mean molecular weight profile $z_I, z_T, \sigma_I$ and $\sigma_T$, (2) the diffusion coefficients, characterized by the Prandtl number Pr and the diffusivity ratio $\tau$, and (3) the system timescales, characterized directly by the global cooling time $\tau_{\rm cool}$, and indirectly by the initial temperature and mean molecular weight stratifications $s$ and $\Delta \mu$. These must therefore be carefully selected in order to achieve dynamics that resemble the ones expected to take place in the WD, while being mindful of computational restrictions on the simulation resolution and total integration time. 

Starting with the diffusion coefficients Pr and $\tau$, it is informative to determine what the latter are in the WD at the core-mantle interface. Characteristic temperatures and densities at this location, just prior to the onset of convection, are
$T \approx \unit[2\times10^8]{K}$ and
$\rho \approx \unit[1 \times 10^7]{\gcc}$ \citep{Brooks2017a}.  We estimate the transport
properties by evaluating existing literature estimates at these
conditions for a 50/50 C/O mixture ($\bar{Z} \approx 7$,
$\bar{A} \approx 14$).
The electrons dominate the thermal conductivity and
$\kappa_T \sim \unit[10^{3}]{cm^2\,s^{-1}}$ \citep{Itoh1983}.
Likewise, the electronic contribution to the shear viscosity dominates
and we have $\nu \sim \unit[10^{-1}]{cm^2\,s^{-1}}$ \citep{Itoh1987}.
Thus the Prandtl number of this material is $\Pr \sim 10^{-4}$.  From
\citet{Beznogov2014}, we get a diffusion coefficient
$\kappa_\mu \sim \unit[4 \times 10^{-3}]{cm^2\,s^{-1}}$.  This implies the
diffusivity ratio (reciprocal Lewis number) is
$\tau \sim 3\times10^{-6}$.

We note that initial size of the interface region is of order a pressure scale height
$H \sim \unit[10^8]{cm}$.  The magnitude of the destabilizing
composition gradient can be expressed as the composition part of the
Brunt-V\"ais\"al\"a frequency (i.e. Equation~\eqref{eq:brunt_B} assuming $\gradT = \grada$)
which is  $N^2 \sim -\unit[0.1]{s^{-2}}$.  This implies that the Rayleigh number is
\begin{equation}
  \label{eq:Ra}
  \Ra = \frac{|N^2| H^4}{\kappa \nu} \sim 10^{29}~,
\end{equation}
where we used the diffusivities from the previous paragraph.
Given this large Rayleigh number, convection is expected to be extremely efficient, which suggests that  temperature should be fully mixed rather than remaining at the radiative gradient.

Unfortunately, such small values of Pr and $\tau$ are not computationally achievable in direct numerical simulations. Instead, we select for simplicity values that are closer to one (but still significantly below one). This will result in numerical simulations that qualitatively reproduce the correct dynamics but cannot be expected to be quantitatively accurate\footnote{Note that since the simulations are 2D quantitative accuracy cannot be achieved anyway.}. In order to quantify the impact of varying the parameters towards their true stellar values, we will compare the results of two sets of parameters:  $\Pr = \tau = 0.1$ and $\Pr = \tau = 0.01$, respectively.  

In the fingering regime, typical nondimensional turbulent structures have size of order 10 \citep{Garaud2018}, while in the convective regime we expect them to be as large as the convective zone. With the selection of Pr and $\tau$ made above, the size of the viscous and compositional boundary layers appearing at the edges of these turbulent structures will be of order 0.1 to 1, so this sets the minimum nondimensional lengthscale that needs to be resolved. The total domain size will therefore be limited by the computational power available given the required resolution. In what follows, we take $L_x = 800$ and $L_z = 1200$. We have verified that the horizontal size is large enough to contain many convective eddies (so horizontally-averaged quantities are meaningful), and that the vertical size is large enough to have negligible influence on the system dynamics. 
With this vertical size, the convective region does not reach the domain boundaries or $z_T$ during the simulation.

The selection of the remaining parameters is guided by the following considerations. We first need to ensure that the cooling time is substantially shorter than the fingering mixing timescale across the core-mantle interface.  If this were not the case, it would contradict the idea that the thermohaline mixing is sufficiently inefficient as to not interfere with the flame \citep{Denissenkov2013b}. In our nondimensional model, the turbulent diffusivity due to fingering convection is equal to  $\tau \num $, where $\num$ is the Nusselt number associated with compositional transport in fingering convection, which has typical value \citep{Garaud2018} of $\num \sim 10$ at the selected values of ${\rm Pr}$ and $\tau$ (except very close to the onset of overturning convection where it could be much larger), so we require
\begin{equation}
  \frac{\sigma_I^2}{10 \tau} \gg \tau_{\rm cool} ~.
\end{equation}
We also have to ensure that the cooling time is much larger than the typical convective growth and turnover time, and this can be verified a posteriori. Note, however, that in the selected unit system the Brunt-V\"ais\"al\"a frequency is proportional to ${\rm Pr}^{1/2}$, so the convective turnover timescale is expected to scale with ${\rm Pr}^{-1/2}$. As such, we must select a larger $\tau_{\rm cool}$ for the lower ${\rm Pr}$ simulations. 

We also need to ensure that the initial condition is stable against overturning convection. The Ledoux criterion for stability, when expressed non-dimensionally at $z = z_I$ (which is the most unstable location in the domain at $t = 0$),  
implies 
\begin{equation}
 \frac{\Delta \mu}{2 \sigma_I} < s + 1~. 
  \end{equation}
Finally, the selected value of $s$ needs to remain of order unity to realistically model the conditions within the WD, and the values of $\sigma_I$ and $\sigma_T$ must be much smaller than $L_z$.  

\begin{figure}
  \includegraphics[width=\columnwidth]{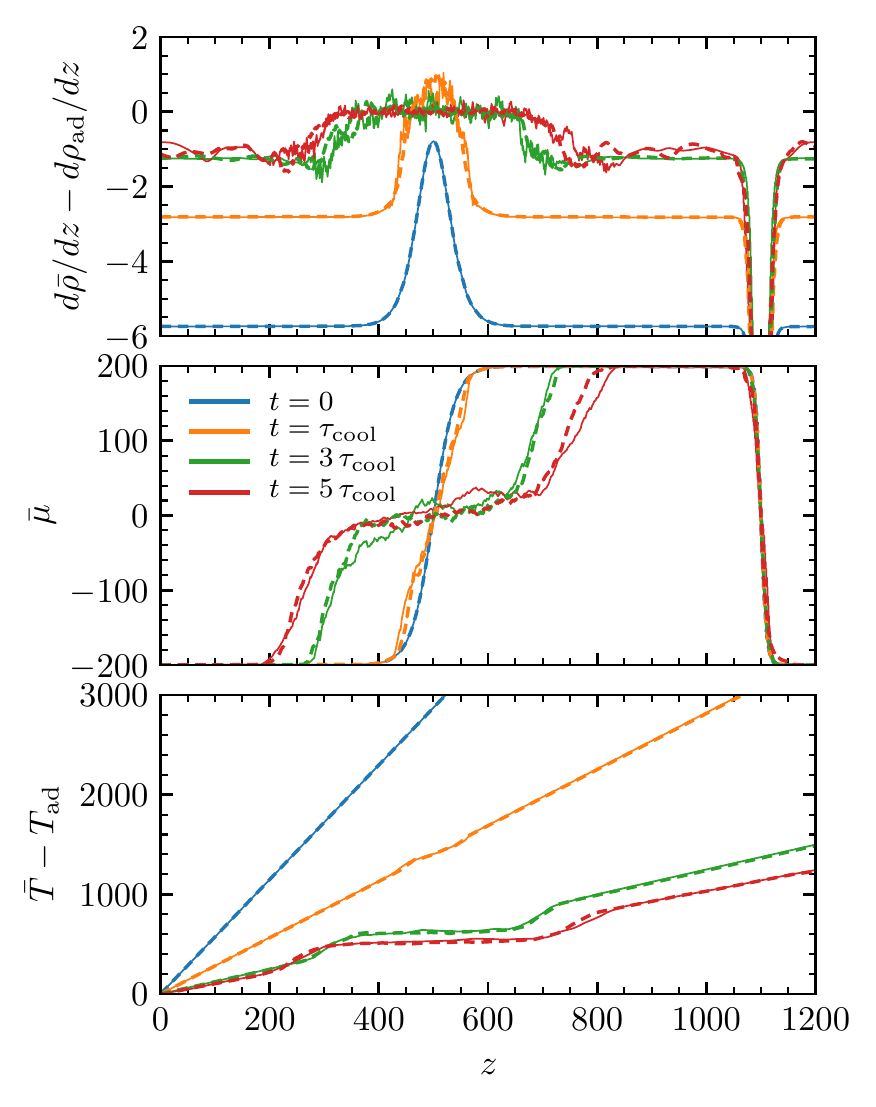}
  \caption{Profiles of the horizontally averaged non-dimensional potential density gradient $d \bar \rho/dz - d\rho_{\rm ad}/dz$, mean molecular weight $\bar \mu$ and potential temperature $\bar T - T_{\rm ad}$ as a function of $z$. Initial profiles (at $t = 0$) are shown along with profiles after one, three and five cooling times.  The solid lines show results for ${\rm Pr} = \tau = 0.01$, $\tau_{\rm cool} = 120$, while the dashed lines show results for ${\rm Pr} = \tau = 0.1$, $\tau_{\rm cool} = 40$.
Note that the domain boundaries, as well as $z_T$, remain at all times far from the convective region.}
  \label{fig:Profiles}
\end{figure}

Based on these considerations, we choose the parameters
$\Delta \mu = 400$, $\sigma_I = 40$, and $s = 5$. In order to ensure that the cooling timescale be much larger than the convective timescale, we pick $\tau_{\rm cool} = 40$ when $\Pr = \tau = 0.1$, and  $\tau_{\rm cool} = 120$ when $\Pr = \tau = 0.01$. The remaining parameters are selected to be $z_I = 500$, $z_T = 1100$, and
$\sigma_T = 10$. The profile $\widehat{\mu}_{\rm init}(z)$, and its corresponding potential density gradient profile $d\widehat{\rho}_{\rm init}/dz - d\rho_{\rm ad}/dz= -(s+1) + d\widehat{\mu}_{\rm init}/dz$ at $t = 0$, are shown in Figure \ref{fig:Profiles}. 

 As required, the potential density gradient is everywhere negative at $t = 0$, so the system is stable to overturning convection (Ledoux-stable). Note that parts of the domain close to $z_I$ are initially unstable to fingering convection, so we expect the fingering instability to first develop in these regions, followed by overturning convection once the stabilizing background temperature gradient has dropped below some critical threshold so that $d\bar{\rho}/dz-d\rho_{\rm ad}/dz$ becomes positive somewhere in the domain.

\subsection{Numerical Results}
\label{sec:paddi} 

\begin{figure*}
  \includegraphics[width=\textwidth]{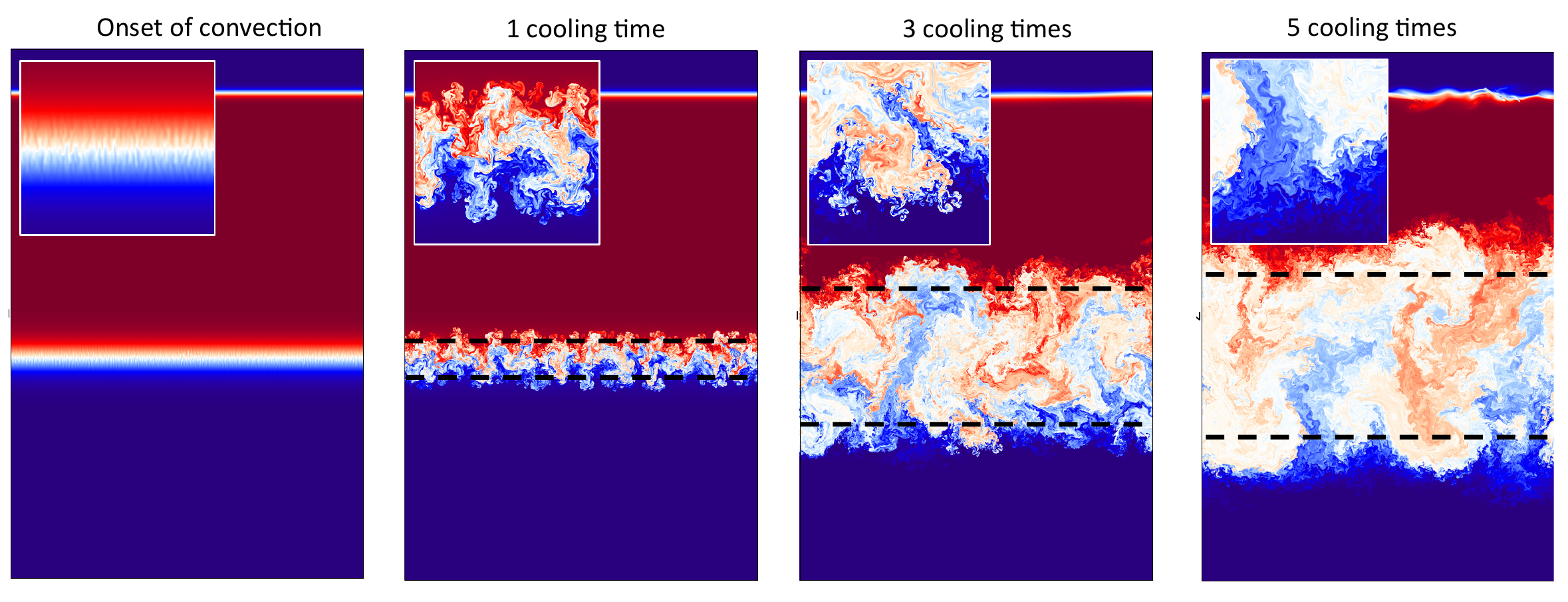}
  \caption{Snapshots of the mean molecular weight profile in the simulation with ${\rm Pr} = \tau = 0.01$, (a) at the onset of overturning convection, (b) after one cooling time, (c) after three cooling times and (d) after five cooling times. The dashed black lines mark the extent of the convective region, where $d \bar \rho /dz - d\rho_{\rm ad}/dz > 0$. In each figure, the inset zooms into an approximately 200$\times$200 region of interest.}
  \label{fig:Snaps}
\end{figure*}

We integrated the set of Equations (\ref{eq:governing-start}-\ref{eq:governing-end}) with the initial conditions described in Section \ref{sec:initial} for 12 cooling times using a resolution of
1024$\times$1536 Fourier modes (equivalently 3072$\times$4608 grid points) for the $\Pr = \tau = 0.1$ run, and for 6 cooling times using 2048$\times$3072 Fourier modes (equivalently 6144$\times$9216 grid points) for the $\Pr = \tau = 0.01$ run. (The latter being much a more computationally demanding case, we were not able to run it for as long relative to the cooling time.) As we shall demonstrate, the qualitative evolution of the system is very similar in both cases when appropriately scaled, showing that the input parameters were correctly selected to be in the desired WD regime (e.g., where the cooling timescale is much faster than the fingering timescale, but much slower than the convective timescale). 
We have verified by inspection of the Fourier spectrum (see Appendix) that all scales of the simulations are fully resolved, from the global convective scale down to the viscous and compositional dissipation scales.

Figure \ref{fig:Snaps} shows representative snapshots of the mean molecular weight perturbation field $\widehat{\mu}$ in the $\Pr = \tau = 0.01$ run, when convection first sets in, and after one, three, and five cooling times. Profiles of the horizontally averaged potential density gradient $d \bar \rho/dz - d\rho_{\rm ad}/dz$, mean molecular weight $\bar \mu$ and potential temperature $\bar T - T_{\rm ad} = \bar T + z$ at these times are shown in Figure \ref{fig:Profiles}. The first snapshot shows the end of the fingering-only phase, and confirms that this instability is too weak to cause much vertical compositional mixing (as expected from the parameters selected) prior to the onset of overturning convection. It does imprint a particular initial size-scale on the developing convective eddies, however. That initial scale rapidly coarsens in the convective phase (second and third snapshots), consistent with the notion that the horizontal size of convective eddies is commensurate with the vertical extent of the convective zone (in the Boussinesq limit at least) which itself grows with time as the stabilizing background temperature gradient decays (see dashed black lines at 1 and 3 cooling times). Once fully developed, the turbulence exhibits a wide range of scales, as expected, from the double-diffusive scale ($l \sim 1$) to the size of the convective region. Inspection of the potential density gradient, potential temperature and mean molecular weight profiles in Figure \ref{fig:Profiles} shows that the convective zone is adiabatic (with $d\bar \rho/dz - 1 \simeq 0$), with substantial ``noise'' due to the fact that the simulations are only two-dimensional. The potential temperature and chemical composition of the convective zone is however not perfectly mixed, and small but finite compensating gradients of both quantities remain. This is interesting from a hydrodynamical point of view and deserves future investigations.

\begin{figure}
  \includegraphics[width=\columnwidth]{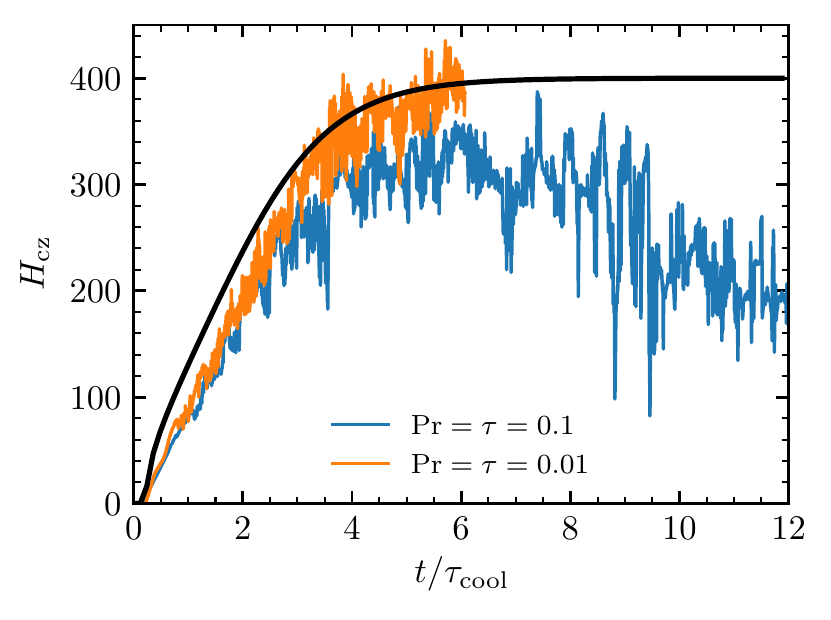}
  \caption{Growth of the size of the convective zone as a function of $t / \tau_{\rm cool}$, for the simulations with $\Pr = \tau = 0.1$ (blue), and $\Pr = \tau = 0.01$ (orange). The black line shows the theoretical predictions obtained by solving Equation~\eqref{eq:hcz}.}
  \label{fig:hcz}.
\end{figure}

The convective region stops growing after about 4 cooling times (with these selected parameters), at which point the background radiative temperature gradient $s e^{-t/\tau_{\rm cool}}$ has become negligible compared with the adiabatic temperature gradient (which is 1 in these units). The size of the convective region at that point is about 370 (in the selected units), and its boundaries are far from the domain boundaries, showing that the reason the convective region stops growing is not because it is  computationally constrained to do so, but instead, because it has reached its equilibrium size. The growth of the size of the convective zone $H_{\rm cz}(t)$ is more easily visualized in Figure \ref{fig:hcz}, together with that of the $\Pr = \tau = 0.1$ simulation. To compute $H_{\rm cz}(t)$, we differentiate the horizontally averaged density profile $\bar \rho(z)$ with respect to $z$, compute the corresponding potential density gradient, then smooth the result using a 50 grid-point wide moving window. We then identify the locations of the points where the potential density gradient first crosses 0 from the bottom and from the top as the base and top of the convection zone respectively. 

We see that the growth of $H_{\rm cz}(t/\tau_{\rm cool})$ is very similar in both runs, confirming that the growth of the convective zone is independent of the properties of convection itself (as long as the convective turnover timescale is short enough compared with $\tau_{\rm cool}$), but instead, only depends on the global structure of the star. In fact, a good estimate for $H_{\rm cz}(t)$ can be obtained by requiring that the total jump in potential density across the convection zone be zero. Since the total jump in potential temperature is $\Delta T_{\rm cz}(t) = H_{\rm cz}(t) \left[ s \exp(-t/\tau_{\rm cool})+ 1\right]$, and the total jump in mean molecular weight is $\Delta \mu_{\rm cz} = \widehat \mu_{\rm init} ( z_I + H_{\rm cz}/2 ) -   \widehat \mu_{\rm init} ( z_I - H_{\rm cz}/2 )$ (assuming that regions outside of the convective zone have not been mixed yet), then $H_{\rm cz}(t)$ can be found by solving the equation
\begin{equation}
H_{\rm cz}(t) \left[ s \exp(-t/\tau_{\rm cool}) + 1\right]  \simeq \Delta \mu \tanh\left( \frac{H_{\rm cz}(t)}{2\sigma_I} \right)~,
\label{eq:hcz}
\end{equation}
using a Newton-Raphson relaxation algorithm. The solution is also shown in Figure \ref{fig:hcz}. We see that this provides a relatively good approximation to the extent of the convective zone measured in the simulations for both simulations. Deviations from the theoretical predictions can be attributed to (1) uncertainties in the measurement of $H_{\rm cz}$ due to the large fluctuations in $d\bar \rho /dz$, and (2) to mixing beyond the edge of the convection zone due to overshoot or fingering, which is ignored in this model. 

\begin{figure}
  \includegraphics[width=\columnwidth]{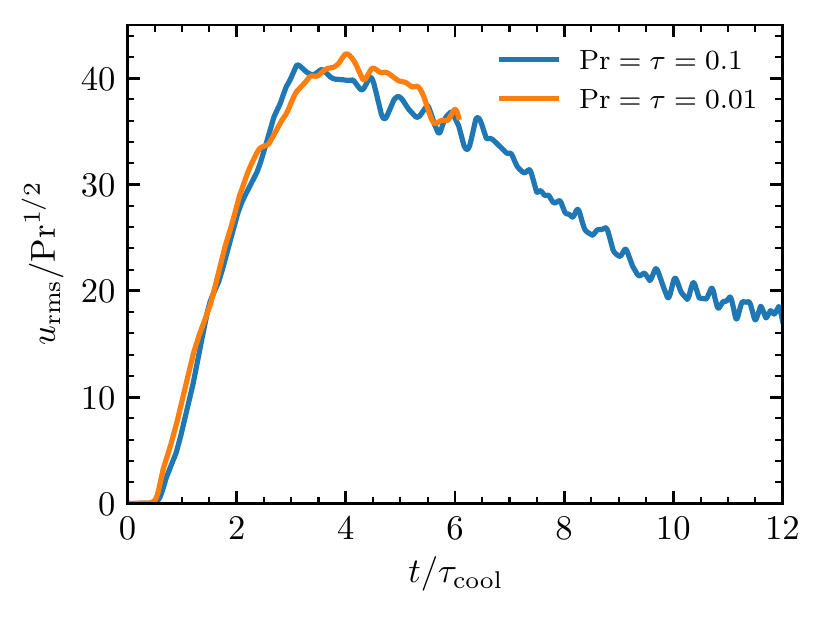}
  \caption{Scaled rms velocity as a function of scaled time for the simulations with $\Pr = \tau = 0.1$ (blue), and $\Pr = \tau = 0.01$ (orange).}
  \label{fig:urms}.
\end{figure}

Figure \ref{fig:urms} shows another more quantitative comparison of the $\Pr = \tau = 0.1$ and  $\Pr = \tau = 0.01$ simulations, by looking at the evolution of the total volume-averaged rms velocity $u_{\rm rms}$ as a function of time. We see that the two are very close to one another when time is scaled by the global cooling time, and when $u_{\rm rms}$ is scaled by the Prandtl number, at least up to $t/\tau_{\rm cool} \simeq 6$. We also see that $u_{\rm rms}$ grows almost linearly with time at early times, mirroring the growth of $H_{\rm cz}(t)$. This confirms that the typical convective velocity in the system is proportional to $H_{\rm cz} N$ where, as discussed earlier, the Brunt-V\"ais\"al\"a frequency $N \propto \Pr^{1/2}$. 

As the convection zone slowly equilibrates, we see in Figure
\ref{fig:Snaps} the development of extended regions on either side
that are mixed by fingering convection, consistent with the
predictions of \citet{Brooks2017a}.  As these regions expand in size,
they mix the areas immediately adjacent to the convection zone and the
latter shrinks in size (see Figure \ref{fig:hcz}, in particular the
$\Pr = \tau = 0.1$ case). This causes the volume-averaged rms velocity
in the convection zone to decrease accordingly (see Figure \ref{fig:urms}). Since
turbulent mixing by fingering convection at $\Pr = \tau = 0.1$ is more
efficient than at $\Pr = \tau = 0.01$, the temperature and
compositional gradients that form on either side of the convection
zone are shallower in the former case, and steeper in the latter (see
Figure \ref{fig:Profiles}). As a result, the decrease in $H_{\rm cz}$
and $u_{\rm rms}$ is faster in the $\Pr = \tau = 0.1$ case than in the
$\Pr = \tau = 0.01$ case. In other words, once cooling is no longer
actively driving the convection, the time-evolution of the system
becomes controlled by fingering-induced mixing at the edge of the
convection zone instead.  A similar transition also occurs
in the \MESA\ models of \citet{Brooks2017a}, as the initially
dominant mixing from overturning convection gradually fades and
thermohaline mixing takes over.

\section{Model comparison}
\label{sec:comparions}

It is difficult to directly compare the results of the \MESA\
calculations from \citet{Brooks2017a} applied to WDs with the
direct numerical simulations of the idealized problem setup in
Section~\ref{sec:numerics}.  However, we can instead compare
predictions for the size and growth of the convection zone made using
the simplified \citet{Brooks2017a} approach, when applied to our
idealized model setup.

\citet{Brooks2017a} assume that convective regions mix composition to
attain neutral stability but that the convection does not affect the
temperature.  Thus in this approach, the imposed time evolution of the
background temperature profile (Equation~\ref{eq:Tgradient}) would be
the true temperature profile.  However, since material in the
convective region is near the neutral stability condition, its
potential density is still near zero, independent of how the
individual temperature and composition gradients behave.

Integrated across the convection zone, the jump in potential density
is then also near zero.  Therefore the \citet{Brooks2017a} approach
also predicts the extent of the convection zone to be given by
Equation~\eqref{eq:hcz}.  In Figure~\ref{fig:hcz} and its surrounding
discussion, we showed that this expression did good job of describing
the time evolution of the size of the convection zone.  Therefore,
while assuming an unmodified temperature gradient fails to faithfully
reproduce the temperature or composition gradients within the
convection zone, it does an equally good job of predicting the
boundary locations.  This suggests that the approach of
\citet{Brooks2017a} accurately predicts which regions of the WD will
be mixed.

By leaving the temperature profile unmodified, the \citet{Brooks2017a}
approach does not capture the energy transport by convection.
Notably, since the background temperature gradient is sub-adiabatic,
the net energy transport from convection will be inward.  During the
early phase where cooling is dominated by neutrino losses, this
additional transport should not affect the cooling timescale.  At
later times, the WD cooling is moderated by the outer layers
\citep{Mestel1952a}, so the details of the deep interior temperature
profile are also unlikely to affect the overall cooling rate.  The
presence of inward energy transport would require a slightly steeper
temperature gradient to achieve the same luminosity.  Since our main
focus is on the chemical structure of the WD, we do not investigate
this effect further.

\section{Conclusions}
\label{sec:conclusions}

We performed direct numerical simulations of an idealized problem that
captures the key features of a cooling hybrid CO/ONe WD.
These simulations were arguably quite idealized: they were two-dimensional, used the Boussinesq approximation for gases \citep{SpiegelVeronis1960}, and assume constant diffusivities, gravity, and coefficients of thermal expansion and compositional contraction. The Boussinesq approximation is only marginally justified, since the interface thickness before the onset of convection in our motivating example (Figure~\ref{fig:mesa-hybrid-composition}) extends for nearly a pressure scale height in the WD. This also implies that the system parameters listed above do vary substantially within the interface in the real star. Finally, two-dimensional dynamics are also known to be problematic at low Prandtl number, often leading to the development of artificial horizontal shear \citep{Goluskin2014, Garaud2015}. However, none of these effects can qualitatively change the results presented here.

Our calculations show that the convective zone is imperfectly mixed in
either composition or entropy, with small but finite compensating
gradients of both quantities remaining.  However, we show that the
size of the convection zone is well-described by a simple expression
that considers only the total change in potential density across the
region.  The treatment of \citet{Brooks2017a} considered only the
effects of compositional mixing and artificially disallowed convection
from changing the temperature gradient.  However, the size of the
mixed region remains driven by the evolution of the background
temperature.  This indicates that despite the approximations, the
overall mixing timescale found by \citet{Brooks2017a} is likely to be
reliable.

Therefore, this study further supports the idea that if massive hybrid
CO/ONe WDs are formed that they will be fully mixed in advance of explosion, and in particular, before carbon burning begins.  When compared to unmixed models,
mixed models seem likely to begin carbon burning at higher densities, as the
mixing reduces the central mass fraction of $^{12}{\rm C}$ and
increases the mass fraction of isotopes that will cool the WD via the
Urca process (i.e., $^{23}{\rm Na}$).  This can affect the explosion
and its nucleosynthesis and so should be taken into account in future
models of these events.

\acknowledgments

Support for this work was provided by NASA through Hubble Fellowship
grant \# HST-HF2-51382.001-A awarded by the Space Telescope Science
Institute, which is operated by the Association of Universities for
Research in Astronomy, Inc., for NASA, under contract NAS5-26555.
P.G. gratefully acknowledges funding by NSF AST-1412951.  The
simulations were run on the Hyades supercomputer at UCSC, purchased
using an NSF MRI grant.  This research made extensive use of NASA's
Astrophysics Data System.


\software{\MESA\ \citep{Paxton2011, Paxton2013, Paxton2015, Paxton2018},
  PADDI \citep{Traxleral2011,Stellmachal2011},
  \texttt{matplotlib} \citep{matplotlib},
  \texttt{ipython/jupyter} \citep{perez2007ipython,kluyver2016jupyter},
  \texttt{NumPy} \citep{numpy},
  \texttt{py\_mesa\_reader} \citep{pmr}
}

\bibliographystyle{aasjournal}
\bibliography{hybridmix}

\appendix

\section{Power Spectra}

\begin{figure}
  \centering
  \includegraphics[width=0.5\textwidth]{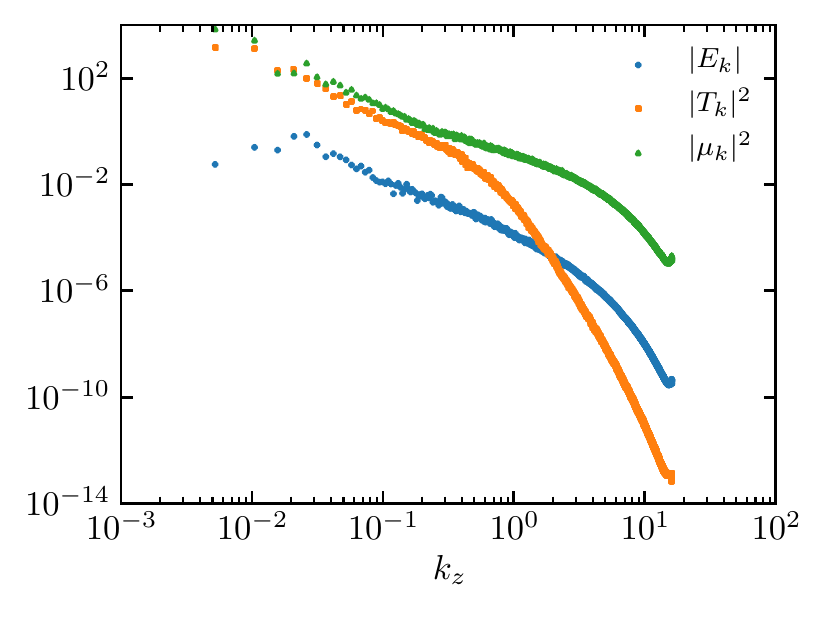}
  \caption{The power spectra of energy, temperature, and mean
    molecular weight in the $\Pr= \tau = 0.01$ simulations at
    $t = 5\,\tau_{\rm cool}$.}
    \label{fig:power-spectra}
\end{figure}

The satisfactory resolution of the code was continually verified for
each simulation. An example is shown in
Figure~\ref{fig:power-spectra}, which presents the power in the
Fourier spectrum of the total kinetic energy, and of the temperature
and compositional perturbations after 5 cooling times in the low $\Pr$
simulation (see rightmost panel of Figure~\ref{fig:Snaps}). This
corresponds to a time where convection is most vigorous (i.e., where
the dissipation scales are the smallest). We see that the code is
resolved well into the dissipation range for each field.

\end{document}